\documentclass[aps,pra,twocolumn,showpacs,preprintnumbers,amsmath,amssymb]{revtex4}

\usepackage{graphicx,subfig}
\usepackage{dcolumn}
\usepackage{bm}

\begin{document}


\title{Enhancement of On-Site Interactions of Tunnelling Ultracold Atoms in Optical Potentials using Radio-Frequency Dressing}

\author{Martin Shotter}
\email{m.shotter@physics.ox.ac.uk}

\author{Dimitrios Trypogeorgos}

\author{Christopher Foot}

\affiliation{Atomic and Laser Physics Department, Clarendon Laboratory, University of Oxford, Parks Road, OX1 3PU, Oxford, UK}

\date{\today}

\begin{abstract}
We show how it is possible to more than double the on-site interaction energy of neutral atoms in optical potentials by the technique of 
radio-frequency (rf) dressing, while maintaining interwell dynamics. We calculate Bose-Hubbard parameters for rf dressed optical lattices 
and arrays of rf dressed dipole traps. We show that decreasing the distance between wells, by the interpolation of wells confining different 
$m_{F}$ states, increases the interaction energy more than decreasing the height of the classically forbidden region between existing wells. 
The schemes we propose have negligible Landau-Zener losses caused by atomic motion; this was a dominant effect in the first 
experimental demonstration of the modification of an optical potential by radio-frequency dressing.
\end{abstract}

\maketitle

The study of complex nonlinear quantum systems is a major area of research in the fields of atomic and condensed matter physics. A 
subject of current interest is the effect of the nonlinear interaction term on the behaviour of ultracold atoms confined in optical lattices. The 
dynamics of ultracold bosonic atoms in such a system has been shown to be described by the Bose-Hubbard Hamiltonian 
\cite{RefWorks:55}
\begin{equation}
H=-J \sum_{<i,j>} a^{\dagger}_{i}a_{j} + \frac{U}{2}\sum_{i} \hat{n}_{i}(\Hat{n}_{i}-1)
\end{equation}
where $i$ and $j$ denote lattice sites of a homogeneous lattice. The ground state of the Bose-Hubbard Hamiltonian passes from superfluid 
to Mott insulator as the parameters controlling $U$ and $J$ are varied; this behaviour has been demonstrated experimentally 
\cite{RefWorks:6}.

The on-site interaction energy, the Hubbard $U$, is the dominant parameter characterising interactions between ultracold atoms in optical 
lattices. As such it plays a pivotal role in phase transitions \cite{RefWorks:6} and entanglement \cite{RefWorks:66}. The ability to generate 
complex entangled states has drawn interest to these systems for the purposes of quantum computing \cite{RefWorks:120} and quantum 
simulation \cite{RefWorks:89}. Typically, the magnitude of $U$ controls the purity of the resulting many-particle state, or the speed at which 
this entanglement may be generated \cite{RefWorks:12}. It is likely that there are a significant number of `impurity' atoms present in the Mott 
insulator states currently being made \cite{RefWorks:66,RefWorks:372}; for applications such as quantum computing and simulation it is 
desirable to reduce the number of these impurities as much as possible. Furthermore, by increasing the on-site interaction relative to the 
tunnelling energy, it may be possible to push optical lattice systems into new regimes, for example where the on-site interaction energy is 
greater than the band gap \cite{RefWorks:242}.

In this paper we study ways to increase the on-site interaction energy, $U$, while still maintaining slow interwell dynamics, using the 
technique of radio-frequency (rf) dressing of optical potentials. We study both optical lattices and arrays of highly focused laser dipole 
spots, both seen as promising systems for quantum information processing \cite{RefWorks:89,RefWorks:376}. We restrict our study to 
dressing the optical potential along a single direction of the lattice. The atoms in these potentials will have far lower Landau-Zener loss rates 
than in the recent experiment \cite{RefWorks:348}, and low collisional loss rates. For a number of specific experimental configurations we 
calculate the enhancement factor for $U$ at a particular interwell tunnelling rate. 

Rf dressing has been used in a variety of experiments to dress magnetic potentials \cite{RefWorks:334, RefWorks:336, RefWorks:375}. 
Dressing of optical potentials has received much less attention, although recently both theoretical \cite{RefWorks:383} and experimental 
\cite{RefWorks:348} studies have been carried out. A dominant feature of the recent experiment \cite{RefWorks:348} was high loss rates, 
attributed to non-adiabatic Landau-Zener transitions \cite{RefWorks:383}. We should mention other methods which may be adapted to 
increase on-site interactions in a lattice include tunable scattering lengths and Feshbach resonances \cite{RefWorks:379,RefWorks:378}, 
and other methods proposed to give sub-half-wavelength structure to optical lattices, for example Raman processes 
\cite{RefWorks:358,RefWorks:361}.

\begin{figure*}[t]
  \centering
  \subfloat[`Truncated' Sinusoidal Potential]{\label{pot11}\includegraphics[width=60mm]{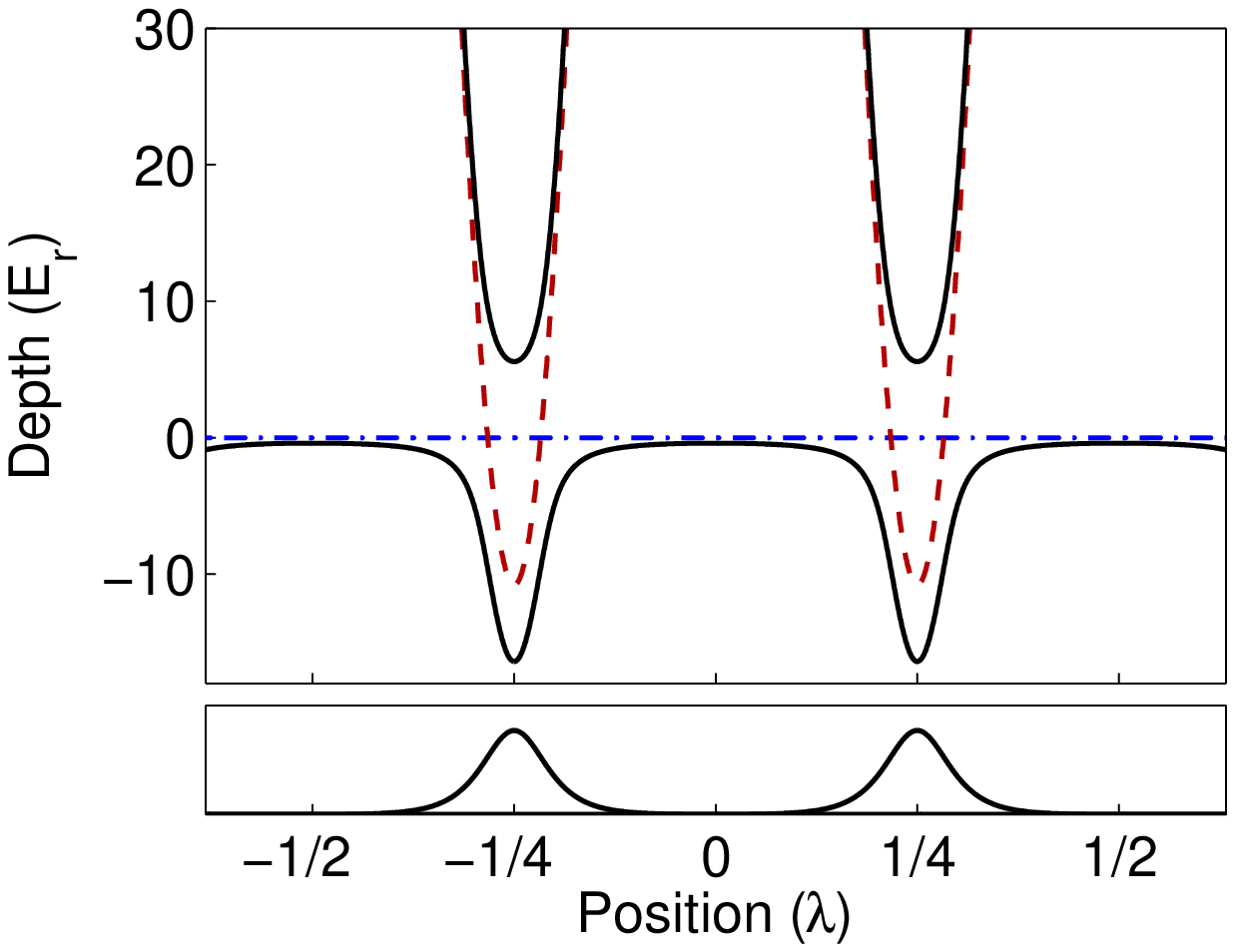}}
  \subfloat[`Interpolated' Sinusoidal Potential]{\label{pot12}\includegraphics[width=56mm]{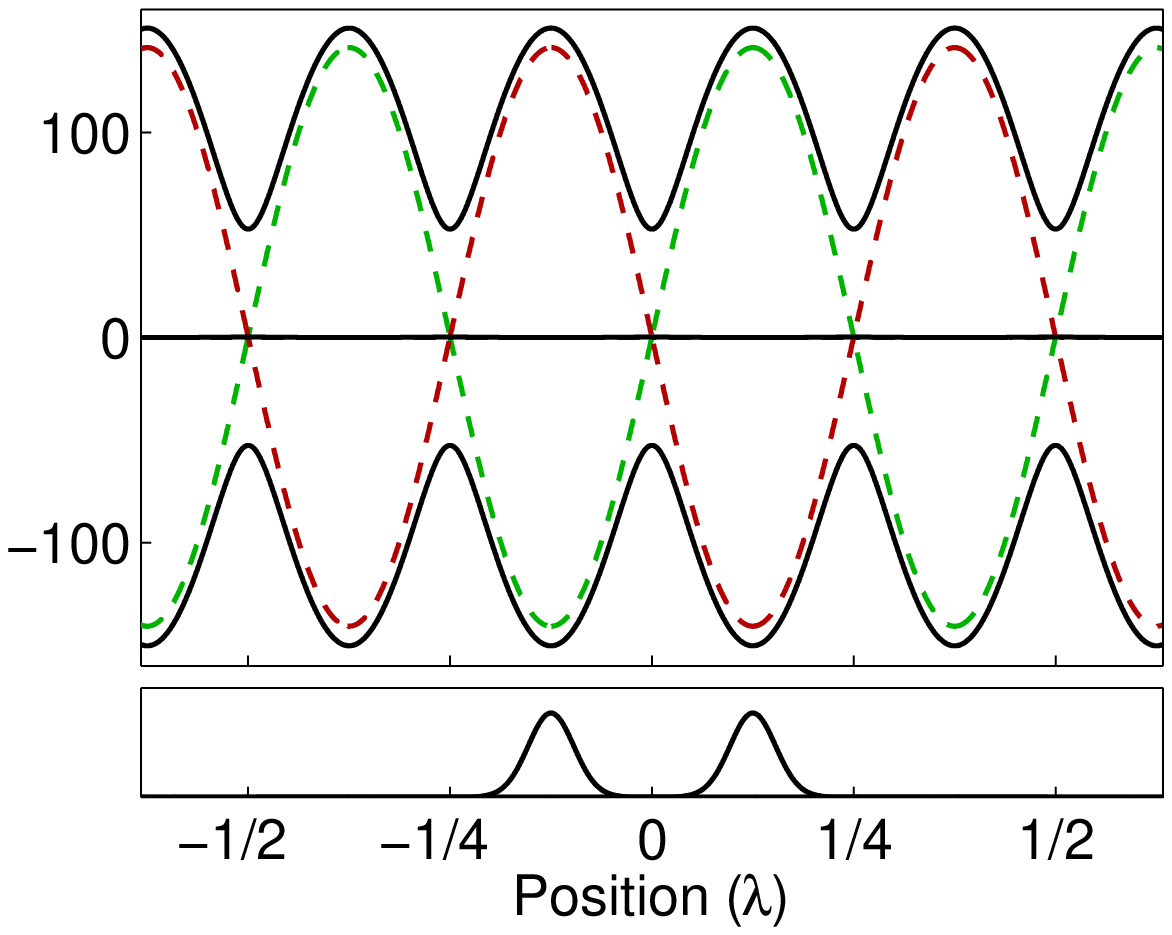}}
  \subfloat[Gaussian Array]{\label{pot13}\includegraphics[width=57mm]{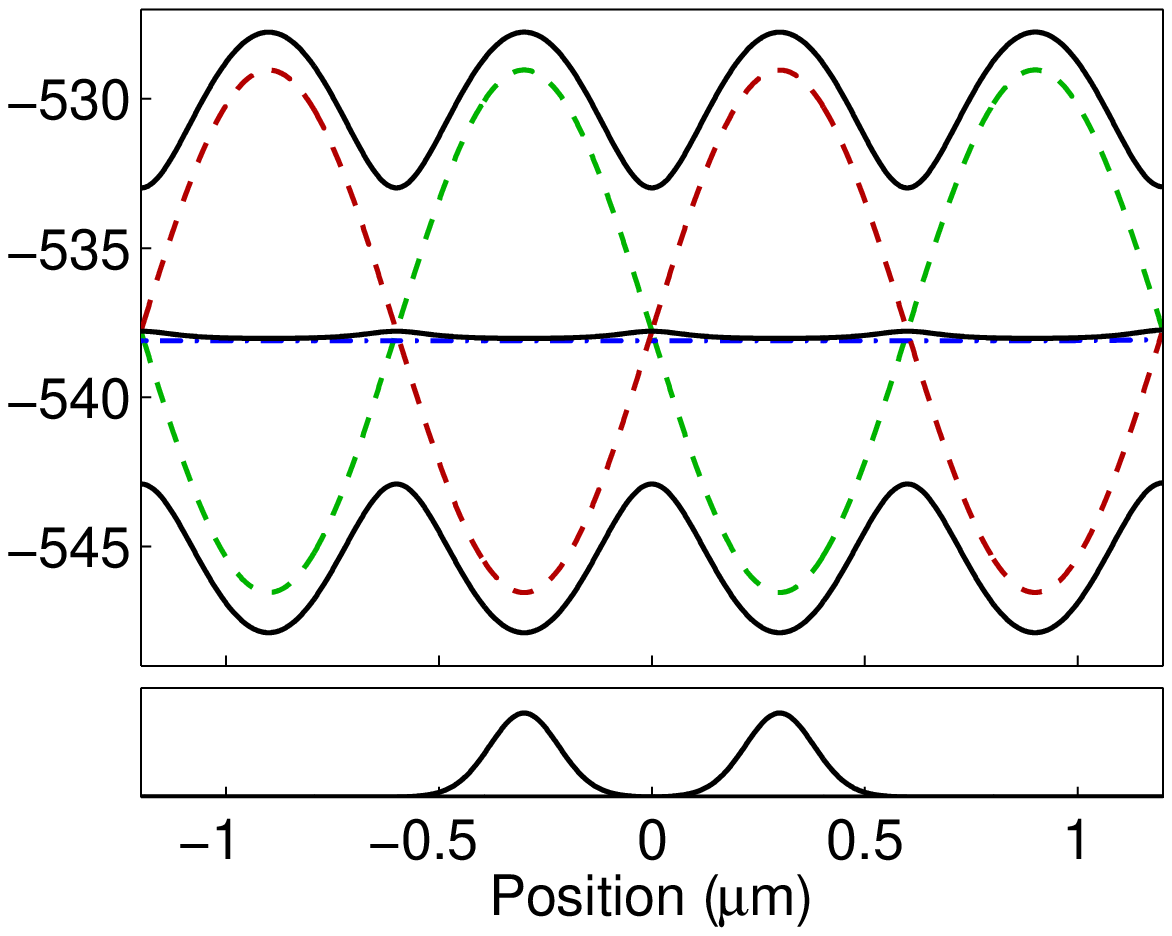}}
  \caption{(Colour online) The types of RF dressed potentials considered in this paper. Dressed potentials are solid black lines, undressed 
potentials are broken lines [$m_{F}=+1/-1$ (green/red dashed), $0$ (blue dash-dotted)]. The Wannier functions for the lowest state are 
displayed below the potentials for two of the wells. The $m_{F}=+1$ state of (a) is at an energy substantially higher than the states shown. 
The optical lattices of (a) and (b) have $\sigma^{+}$ polarisation. The recoil energy is defined as $E_{r}= h^{2} / 2 m  \lambda^{2}$. 
Parameters: \protect\linebreak (a) $\lambda=790.06\,$nm, $B_{0}=300\,$G, $\nu_{rf}=213.65\,$MHz, $\Omega_{rf}/2\pi= 70\,$kHz, 
$P=32\,$mW, $w_{0}=50\,\mu$m, $\nu_{\perp 1,2}=30\,$kHz \protect\linebreak (b) $\lambda=790.06\,$nm, $B_{0}=4\,$G, 
$\nu_{rf}=3.295\,$MHz, $\Omega_{rf}/2\pi= 270\,$kHz, $P=35\,$mW, $w_{0}=50\,\mu$m, $\nu_{\perp 1,2}=30\,$kHz \protect\linebreak (c) 
$\lambda=800\,$nm, $B_{0}=4\,$G, $\nu_{rf}=2.78\,$MHz, $\Omega_{rf}/2\pi=25\,$kHz, $P=25\,\mu$W, $w_{0}=1\,\mu$m, $\nu_{\perp 
1,2}=30\,$kHz, $\Delta x = 0.6w_{0}$ } 
  \label{gr1}
\end{figure*}

The phase transition from the superfluid state ceases to be adiabatic when the inverse timescale $1/ \tau$ becomes of the same order of 
magnitude as the frequencies of the lowest lying excitations, which are $\sim U$ in the limit $U \gg J$; increasing $U$ will decrease the 
number fluctuations in the final state. Furthermore, when finite temperature effects are taken into account, it has been shown that increasing 
$U$ increases the purity of the final state \cite{RefWorks:371}, as one would expect from thermodynamic considerations. Both temperature 
and non-adiabatic defects are exponentially suppressed by increasing $U$, as discussed below. 

We treat the case with the spin dependent potential only along one direction of the lattice; dressing a spin dependent potential of more than 
one dimension tends to lead to structures with extended potential minima \cite{RefWorks:383}. The methods we describe may be used with 
or without additional tunnelling in the perpendicular directions (along spin independent potentials). Note that a low dimensional lattice can 
already have a larger $U$ than a higher dimensional lattice due to tight confinement being possible along the perpendicular directions. 

In this work we confine our analysis to bosons, specifically to $^{87}$Rb. We choose to work in the $F=1$ lower hyperfine state with three 
magnetic substates. With $B_{rf}$ and $B_{0}$ perpendicular ($B_{0}$ being the static magnetic field), there will be equal intensities of 
$\sigma^{+}$ and $\sigma^{-}$ rf dressing fields. We assume a state independent ($\pi$ polarised, or with large frequency detuning) optical 
lattice in the $y$ and $z$ directions, meaning that the difference in energy between magnetic substates is independent of $y$ and $z$. We 
can then write, for these three substates,
\begin{equation}
H=\left( \begin{array}{ccc}
V_{+1}(\mathbf{x})+\delta_{+1}&\Omega_{rf} / 2&0\\
\Omega_{rf} / 2&V_{0}(\mathbf{x})&\Omega_{rf} / 2\\
0&\Omega_{rf} / 2&V_{-1}(\mathbf{x})+\delta_{-1}
\label{matr} 
\end{array}
\right)
\end{equation}
with $\delta_{ \pm 1}=  \Delta E_{\pm 1, 0} / \hbar \pm \omega_{rf} $. The optical dipole potentials $V_{mF}(\mathbf{r})$ can be shown to be 
\cite{RefWorks:381} 
\begin{equation}
V_{m_{F}}(\mathbf{r})=\frac{\pi c^{2} \Gamma} {2 \omega^{3}} \left( \frac{ 1 - P g_{F} m_{F} } {\Delta_{D_{1}}} + \frac{ 2 + P g_{F} m_{F} } 
{\Delta_{D_{2}}} \right) I(\mathbf{r})
\end{equation} 
where $\Delta_{D_{1}}$ and $\Delta_{D_{2}}$ are the frequency differences from the $D$ lines, 
$\Gamma=(\Gamma_{D_{1}}+\Gamma_{D_{2}})/2$, $\omega=(\omega_{D_{1}}+2 \omega_{D_{2}})/3$ and $P= \pm 1, 0$ for optical 
$\sigma^{\pm}, \pi$ polarisation.

Using a Born-Oppenheimer-type approximation \cite{RefWorks:377}, the internal and external degrees of freedom of the atom can be 
decoupled as long as the kinetic energy of the atoms is much less than the energy spacing of the dressed levels. This is the case for typical 
$\Omega_{rf}$, although Landau-Zener losses may occur as a result of this condition being weakly violated, as discussed below. Under this 
approximation, the eigenvalues of the above matrix give the potential energy of the atoms in the corresponding dressed eigenstate. The rf 
coupling between the magnetic substates gives rise to avoided crossings, as shown in Fig.\ \ref{gr1}.

\begin{figure*}[t]
  \centering
  \subfloat[`Truncated' Sinusoidal Potential]{\label{g1}\includegraphics[width=61mm]{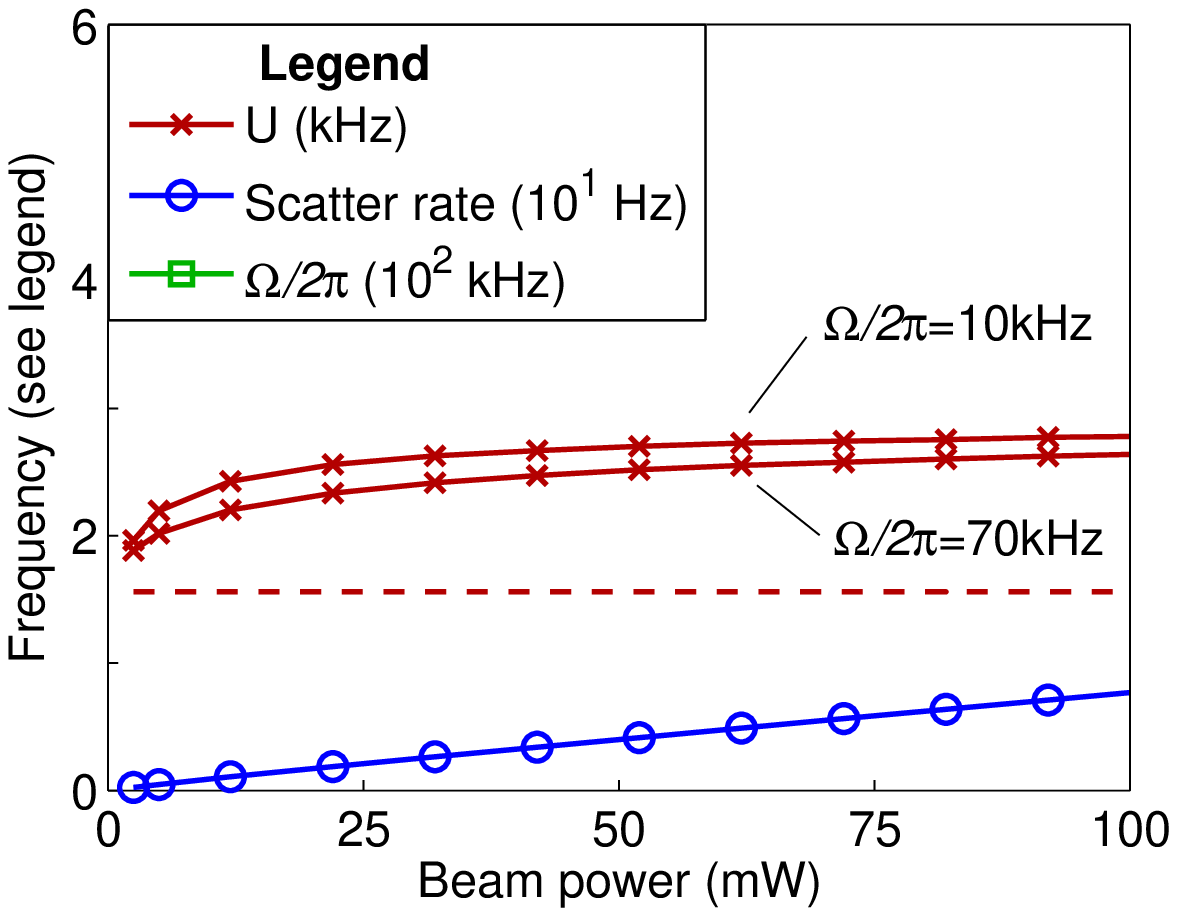}}
  \subfloat[`Interpolated' Sinusoidal Potential]{\label{g2}\includegraphics[width=57mm]{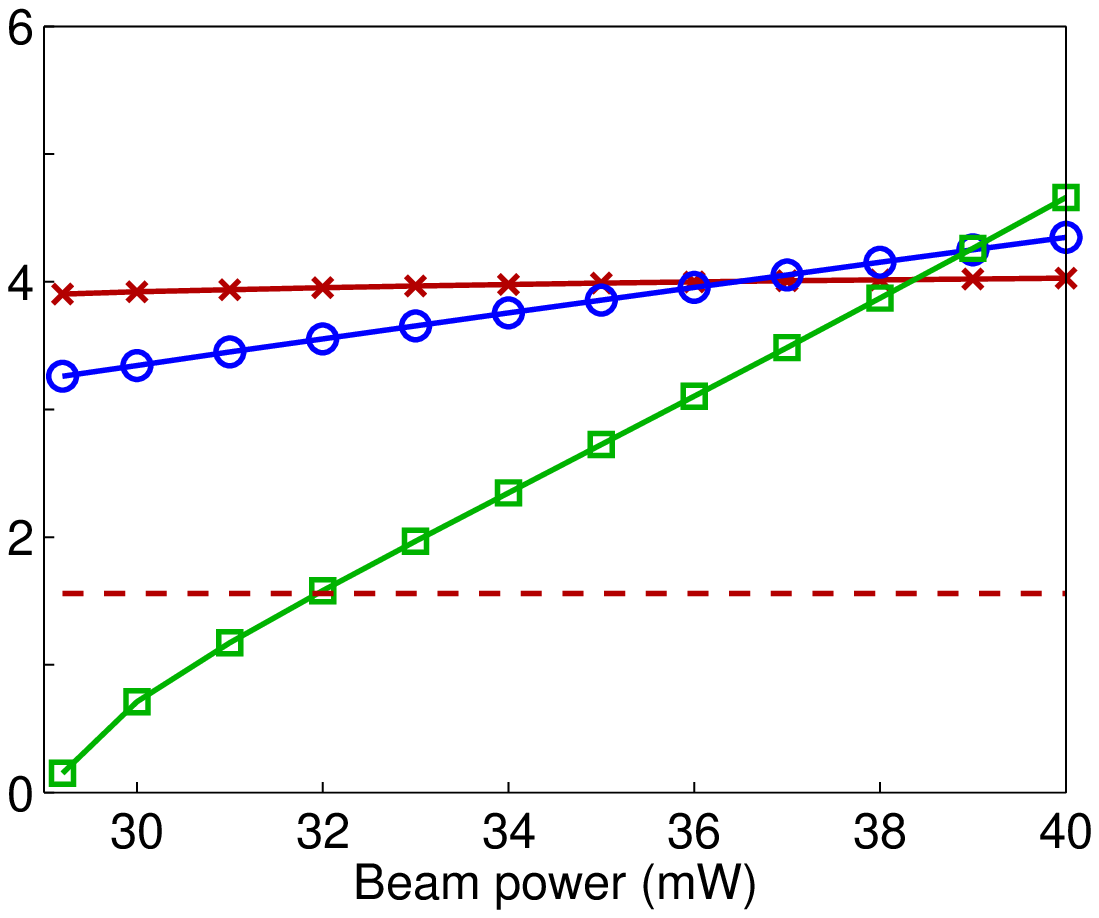}}
  \subfloat[Gaussian Array]{\label{g3}\includegraphics[width=57mm]{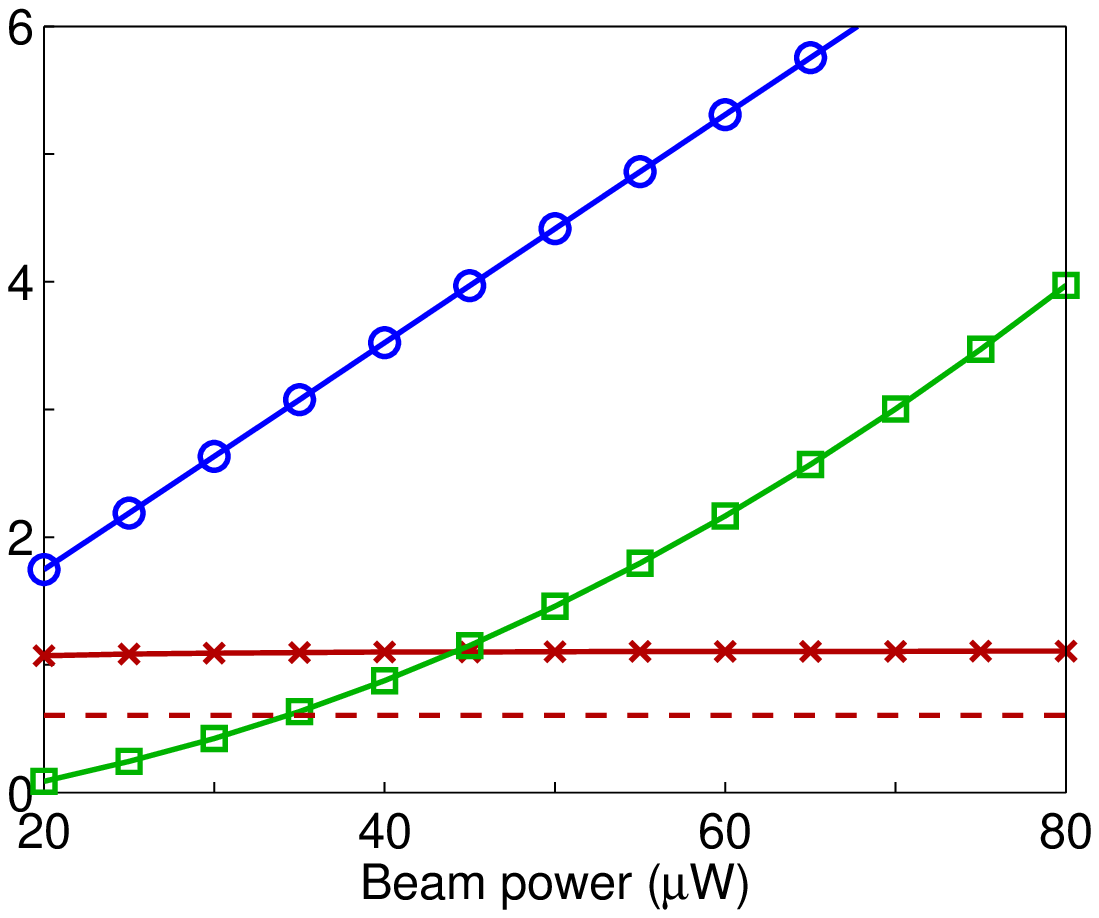}}
  \caption{(Colour online) Calculated properties of atoms in the ground state of the dressed potentials. The optical power $P$ is the 
independent variable; dressing parameters $\nu_{rf}$ and $\Omega_{rf}$ are chosen subject to the conditions that the neighbouring wells 
are the same depth and that the tunnelling parameter $J/h$ is $25\,$Hz. These conditions specify $\nu_{rf}$ and $\Omega_{rf}$ for (b) and 
(c). For (a) the conditions only specify $\nu_{rf}$, so $\Omega_{rf}$ can be chosen arbitrarily; two choices are shown. The dashed horizontal 
line gives the undressed $U$ for comparison, a single value calculated for $J/h=25\,$Hz in the undressed $\sigma^{+}$ or $\sigma^{-}$ 
potential. The scattering rates are from the dressed lattice beams only. Aside from $P$, $\nu_{rf}$ and $\Omega_{rf}$, the parameters of 
the potentials are as given in Fig.\ \ref{gr1}.} 
  \label{gr2}
\end{figure*}

We calculate the one dimensional Wannier functions for the lowest eigenstate of the Hamiltonian (Eqn. \ref{matr}), corresponding to the 
local ground state of the atoms. We diagonalise the Hamiltonian of a single potential period, and so find the eigenfunctions of the dressed 
potential by using Bloch's theorem. We find the maximally localised Wannier functions at the lattice sites by recursively rephasing the 
eigenstates \cite{RefWorks:352} before summing over quasimomentum. From the lowest band Wannier functions we calculate the 
Bose-Hubbard parameters
\begin{eqnarray}
U=2a_{s} \hbar \sqrt{\omega_{\perp 1} \omega_{\perp 2}} \int |w_{0}(x)|^{4} dx \label{U} \\
J=\frac{1}{N_{q}} \sum_{q} E_{q} e^{iqx_{r}}
\end{eqnarray}
with $\omega_{\perp 1,2}$ the trapping frequencies in the perpendicular directions, $q$ the quasimomentum, $E_{q}$ the energy of the 
eigenstate with quasimomentum $q$,  $x_{r}$ the separation between neighbouring lattice sites, and $a_{s}$ the scattering length in the 
dressed state (for $^{87}$Rb, approximately equal to the undressed scattering length as $a_{s}\approx a_{t}$ \cite{RefWorks:380}). The 
photon scattering rate is calculated from the Wannier functions and details of the dressed potential.  

We compare the on-site interaction energy between dressed and undressed potentials for a certain value of Hubbard $J$. We choose 
$J/h=J_{lim}/h=25\,$Hz so that the tunnelling time $\tau_{tun}=h/2zJ$ is $\tau_{tun}=10\,$ms ($z$ being the coordination number). We 
choose this value as this is around the limit where slow interwell dynamics still occur on a typical experimental timescale; as $J$ is 
decreased further the phase transition will go from being an adiabatic to a non-adiabatic process. Interwell dynamics then cease, with the 
final purity of the Mott insulator depending on $U_{lim}$, the value of $U$ when $J=J_{lim}$.

The first potential we consider is given in Fig.\ \ref{pot11}, the `truncated sinusoidal' potential. This is an optical lattice at the $^{87}$Rb, 
$F=1$, $m_{F}=0$ tune-out wavelength of 790.06nm \cite{RefWorks:320}. The magnetic field magnitude and direction are chosen so that 
the $m_{F}=+1$ state is detuned well above the other two states by the nonlinear Zeeman effect. Experimentally plausible parameters are 
chosen. The lowest parts of the $m_{F}=-1$ potential intersect the flat $m_{F}=0$ potential. Once dressed, the lowest adiabatic potential 
differs greatly from a sinusoidal shape, with `pockets' of strong confinement separated by an almost flat potential. There is typically only a 
single bound state at each site. The enhancement of $U_{lim}$, for the chosen value of $J_{lim}$, in this potential is given in Fig.\ \ref{g1} 
for $\Omega_{rf}=2\pi \, \times \, 70\,$kHz and $2\pi\,\times\,10\,$kHz; the maximum increases we find are around 70\% and 80\% 
respectively. The enhancement factor is only weakly dependent on $\Omega_{rf}$ for a large range of rf power.

The second potential we consider is given in Fig.\ \ref{pot12}, the `interpolated sinusoidal' potential. Optically, it is very similar to the 
truncated sinusoid, but at a much lower magnetic field of a few Gauss, so the $m_{F}=\pm1$ states are approximately symmetrical around 
the $m_{F}=0$ state. In this case the rf dressing doubles the number of wells in the lattice. The enhancement of $U_{lim}$ in this potential is 
given in Fig.\ \ref{g2}; the typical increase is around 150\%.

The last dressed potential (Fig.\ \ref{pot13}) we consider is based on an array of independently addressable dipole traps\footnote{To use 
Wannier functions we assume that the region of interest is far enough away from the edge of the array that we can neglect the finite array 
size.}. We choose neighbouring dipole traps to have the same intensity but opposite $\sigma$ polarisation. Due to the rf dressing, the 
barrier between the potential minima can be considerable even when neighbouring Gaussian spots are optically unresolvable. The 
enhancement of $U_{lim}$ is shown in Fig.\ \ref{g3}; the maximum enhancement due to the dressing is around 80\%. Note that, for 
potentials (b) and (c), the atomic spin adiabatically flips when a single atom tunnels from one site to its neighbour, enabling sublattice 
addressability and readout.

The results of these calculations show that the dressed $U_{lim}$ is largely independent of the optical power, but is dependent on the 
dressing scheme. The results for the `truncated' sinusoidal potential show that radically altering the shape of the periodic potential, to 
confine tighter while still allowing tunnelling, but without changing the distance between wells, can only increase $U_{lim}$ by a modest 
amount. However, $U_{lim}$ becomes substantially larger if atoms are confined in the `interpolated' sinusoidal potential, with neighbouring 
wells separated by $\lambda/4$. The interpolation method would therefore seem the most promising way to increase $U_{lim}$ by rf 
dressing an optical lattice. 

The results show that the on-site interaction energy of atoms confined in a focussed Gaussian array can also be increased by the 
interpolation technique. The typical values of $U_{lim}$ in this case are not greatly less, for our chosen parameters, than for atoms in an 
undressed 3D counterpropagating lattice. This raises the prospect of site-addressable highly number squeezed states in a hybrid focussed 
Gaussian / 2D optical lattice apparatus, with a comparable on-site interaction energy per atom to that observed in a Mott Insulator in a 3D 
optical lattice \cite{RefWorks:6}.

Even a modest increase in $U_{lim}$ could be useful, though, as the number of imperfections is likely to be a strong function of $U_{lim}$. 
Estimates of the thermal excitations would be $n_{th}\approx Ae^{\frac{-U}{kT}}$ (the $J\ll U,kT\ll U$ limit of the Bose-Einstein distribution) 
and of non-adiabatic excitations $n_{na}\approx Be^{-CU^{2}}$; the latter from the form of the Landau-Zener avoided crossing transition 
probability \cite{RefWorks:384} in the limit $J\ll U \approx \epsilon_{01}$, where $\epsilon_{01}$ is the energy difference between the 
ground and first excited states of the many body system. 

We briefly consider how atoms may be loaded into the potentials (when $J > U$). Adiabatic loading of the `truncated' sinusoidal potential 
can be seen to be  accomplished by simply ramping the rf frequency, once atoms are trapped in the bare lattice. Adiabatic loading of the 
`interpolated' sinusoidal potential and Gaussian array may be accomplished by ramping the laser intensity (with the rf already on), but the 
mechanism is less obvious. In effect, during the final stages of the ramping of the lattice intensities, the atoms adiabatically delocalise 
between the two sets of potential minima as long as the rate of change of the offset between the two sublattices $\epsilon_{i}$ is much less 
than the other Hamiltonian parameters.

It is important to note that spin-changing collisions and other 2-body collisional loss rates are greatly suppressed for atoms trapped in the 
lowest eigenstate \cite{RefWorks:333,RefWorks:337,RefWorks:334}. This is because a collision which changes the internal states of these 
atoms needs an input of energy which is significantly greater than the energy available from kinetic or potential energy (of order $J$ or $U$). 
The other loss mechanism is Landau-Zener losses arising from atomic motion at the avoided crossings, which have dominated the only rf 
dressed optical potential experiment carried out to date \cite{RefWorks:348}. However, Landau-Zener losses become negligible when the 
atoms are trapped in the lowest eigenstate, as the transition from the lower to the upper dressed level needs energy which is required to 
come from the kinetic energy of the atoms; this kinetic energy $\sim$$J$ is typically very much less than $\Omega_{rf}$, so Landau-Zener 
losses can be expected to be negligible for atoms in the lowest trapped state \cite{RefWorks:385}. 

A remaining loss mechanism, specific to the truncated lattice, is the coupling of weakly bound states to the continuum. This can only occur 
when the highest energy state in the lowest band becomes greater than the binding energy of the site. When $n_{i}U$ and $J$ are 
significantly less than the binding energy of the site, as is the case in the examples given above (assuming around one atom per site), this 
loss mechanism is, to a large extent, suppressed; however with higher numbers of atoms per site there is a chance that some atoms will be 
ejected from the lattice. 

The calculations shown do not specify whether tunnelling is proceeding in the perpendicular (spin-independent) directions; $J$ along the 
dressed lattice direction is independent of $J$ in the other directions. The lattice parameters in the other directions only influence $U$ 
through a multiplicative factor in Eqn. \ref{U}. Therefore, when there is tunnelling in all directions in such a lattice, the enhancement factor for 
$U$ arising from dressing along a single direction will be the same whether or not there is also tunnelling in the perpendicular directions. 
Thus this technique may be used in a 3D lattice (composed of 1D spin-dependent and 2D spin-independent lattices), to enhance $U_{lim}$ 
by the same factor as in purely 1D lattices. 

If spin-dependent potentials are present along more than one direction, the 2D or 3D rf dressed potential cannot be simply expressed as a 
sum of 1D potentials; one consequence is that the potential minima may occur along lines or surfaces rather than at points. A short 
discussion may be found in Ref. \cite{RefWorks:383}; we do not consider these cases further in this paper. 

A practical complication with this scheme, in common with other rf dressed optical potential schemes, lies in the use of magnetic field 
sensitive transitions. The typical magnetic field drift in a laboratory environment is of order $\sim$$1$mG \cite{RefWorks:446}; this would 
manifest as a sublattice-dependent energy offset of order $\sim$$1$kHz. Clearly we need this site-dependent offset to be less than $U$; 
practically this would mean using a magnetic shielding technique such as mu-metal cladding, which can decrease these ambient fields by a 
factor of around 100 \cite{RefWorks:445}.

In conclusion, we show that substantial enhancement of on-site interactions may be achieved by rf dressing optical potentials. We study 
three cases of interest, and show that it is possible to enhance nonlinear parameters by more than a factor of 2, with negligible 
Landau-Zener losses, which should greatly improve the purity of the resulting Mott insulator state. We find that decreasing the distance 
between neighbouring wells has a greater effect on the limiting on-site interaction than modifying the form of the potential between existing 
wells. In summary, the techniques described have the potential to make complex quantum states with neutral atoms purer and faster.

The authors would like to thank A. Daley for a useful discussion, and acknowledge funding from the EPSRC (M.S.), Christ Church, Oxford 
(M.S.), and QIPEST (D.T.).

\bibliography{rf_dressing}

\begin{thebibliography}{30}
\expandafter\ifx\csname natexlab\endcsname\relax\def\natexlab#1{#1}\fi
\expandafter\ifx\csname bibnamefont\endcsname\relax
  \def\bibnamefont#1{#1}\fi
\expandafter\ifx\csname bibfnamefont\endcsname\relax
  \def\bibfnamefont#1{#1}\fi
\expandafter\ifx\csname citenamefont\endcsname\relax
  \def\citenamefont#1{#1}\fi
\expandafter\ifx\csname url\endcsname\relax
  \def\url#1{\texttt{#1}}\fi
\expandafter\ifx\csname urlprefix\endcsname\relax\def\urlprefix{URL }\fi
\providecommand{\bibinfo}[2]{#2}
\providecommand{\eprint}[2][]{\url{#2}}

\bibitem[{\citenamefont{Jaksch et~al.}(1998)\citenamefont{Jaksch, Bruder,
  Cirac, Gardiner, and Zoller}}]{RefWorks:55}
\bibinfo{author}{\bibfnamefont{D.}~\bibnamefont{Jaksch}},
  \bibinfo{author}{\bibfnamefont{C.}~\bibnamefont{Bruder}},
  \bibinfo{author}{\bibfnamefont{J.~I.} \bibnamefont{Cirac}},
  \bibinfo{author}{\bibfnamefont{C.~W.} \bibnamefont{Gardiner}},
  \bibnamefont{and} \bibinfo{author}{\bibfnamefont{P.}~\bibnamefont{Zoller}},
  \bibinfo{journal}{Physical Review Letters} \textbf{\bibinfo{volume}{81}},
  \bibinfo{pages}{3108} (\bibinfo{year}{1998}).

\bibitem[{\citenamefont{Greiner et~al.}(2002)\citenamefont{Greiner, Mandel,
  Esslinger, Hänsch, and Bloch}}]{RefWorks:6}
\bibinfo{author}{\bibfnamefont{M.}~\bibnamefont{Greiner}},
  \bibinfo{author}{\bibfnamefont{O.}~\bibnamefont{Mandel}},
  \bibinfo{author}{\bibfnamefont{T.}~\bibnamefont{Esslinger}},
  \bibinfo{author}{\bibfnamefont{T.~W.} \bibnamefont{Hänsch}},
  \bibnamefont{and} \bibinfo{author}{\bibfnamefont{I.}~\bibnamefont{Bloch}},
  \bibinfo{journal}{Nature} \textbf{\bibinfo{volume}{415}}, \bibinfo{pages}{39}
  (\bibinfo{year}{2002}).

\bibitem[{\citenamefont{Mandel et~al.}(2003)\citenamefont{Mandel, Greiner,
  Widera, Rom, Hänsch, and Bloch}}]{RefWorks:66}
\bibinfo{author}{\bibfnamefont{O.}~\bibnamefont{Mandel}},
  \bibinfo{author}{\bibfnamefont{M.}~\bibnamefont{Greiner}},
  \bibinfo{author}{\bibfnamefont{A.}~\bibnamefont{Widera}},
  \bibinfo{author}{\bibfnamefont{T.}~\bibnamefont{Rom}},
  \bibinfo{author}{\bibfnamefont{T.~W.} \bibnamefont{Hänsch}},
  \bibnamefont{and} \bibinfo{author}{\bibfnamefont{I.}~\bibnamefont{Bloch}},
  \bibinfo{journal}{Nature} \textbf{\bibinfo{volume}{425}},
  \bibinfo{pages}{937} (\bibinfo{year}{2003}).

\bibitem[{\citenamefont{Pachos and Knight}(2003)}]{RefWorks:120}
\bibinfo{author}{\bibfnamefont{J.~K.} \bibnamefont{Pachos}} \bibnamefont{and}
  \bibinfo{author}{\bibfnamefont{P.~L.} \bibnamefont{Knight}},
  \bibinfo{journal}{Physical Review Letters} \textbf{\bibinfo{volume}{91}},
  \bibinfo{pages}{107902} (\bibinfo{year}{2003}).

\bibitem[{\citenamefont{Jaksch and Zoller}(2005)}]{RefWorks:89}
\bibinfo{author}{\bibfnamefont{D.}~\bibnamefont{Jaksch}} \bibnamefont{and}
  \bibinfo{author}{\bibfnamefont{P.}~\bibnamefont{Zoller}},
  \bibinfo{journal}{Annals of Physics} \textbf{\bibinfo{volume}{315}},
  \bibinfo{pages}{52} (\bibinfo{year}{2005}).

\bibitem[{\citenamefont{Jaksch et~al.}(1999)\citenamefont{Jaksch, Briegel,
  Cirac, Gardiner, and Zoller}}]{RefWorks:12}
\bibinfo{author}{\bibfnamefont{D.}~\bibnamefont{Jaksch}},
  \bibinfo{author}{\bibfnamefont{H.~J.} \bibnamefont{Briegel}},
  \bibinfo{author}{\bibfnamefont{J.~I.} \bibnamefont{Cirac}},
  \bibinfo{author}{\bibfnamefont{C.~W.} \bibnamefont{Gardiner}},
  \bibnamefont{and} \bibinfo{author}{\bibfnamefont{P.}~\bibnamefont{Zoller}},
  \bibinfo{journal}{Physical Review Letters} \textbf{\bibinfo{volume}{82}},
  \bibinfo{pages}{1975} (\bibinfo{year}{1999}).

\bibitem[{\citenamefont{Reischl et~al.}(2005)\citenamefont{Reischl, Schmidt,
  and Uhrig}}]{RefWorks:372}
\bibinfo{author}{\bibfnamefont{A.}~\bibnamefont{Reischl}},
  \bibinfo{author}{\bibfnamefont{K.~P.} \bibnamefont{Schmidt}},
  \bibnamefont{and} \bibinfo{author}{\bibfnamefont{G.~S.} \bibnamefont{Uhrig}},
  \bibinfo{journal}{Physical Review A} \textbf{\bibinfo{volume}{72}},
  \bibinfo{pages}{063609} (\bibinfo{year}{2005}).

\bibitem[{\citenamefont{Morsch and Oberthaler}(2006)}]{RefWorks:242}
\bibinfo{author}{\bibfnamefont{O.}~\bibnamefont{Morsch}} \bibnamefont{and}
  \bibinfo{author}{\bibfnamefont{M.}~\bibnamefont{Oberthaler}},
  \bibinfo{journal}{Reviews of Modern Physics} \textbf{\bibinfo{volume}{78}},
  \bibinfo{pages}{179} (\bibinfo{year}{2006}).

\bibitem[{\citenamefont{Frese et~al.}(2000)\citenamefont{Frese, Ueberholz,
  Kuhr, Alt, Schrader, Gomer, and Meschede}}]{RefWorks:376}
\bibinfo{author}{\bibfnamefont{D.}~\bibnamefont{Frese}},
  \bibinfo{author}{\bibfnamefont{B.}~\bibnamefont{Ueberholz}},
  \bibinfo{author}{\bibfnamefont{S.}~\bibnamefont{Kuhr}},
  \bibinfo{author}{\bibfnamefont{W.}~\bibnamefont{Alt}},
  \bibinfo{author}{\bibfnamefont{D.}~\bibnamefont{Schrader}},
  \bibinfo{author}{\bibfnamefont{V.}~\bibnamefont{Gomer}}, \bibnamefont{and}
  \bibinfo{author}{\bibfnamefont{D.}~\bibnamefont{Meschede}},
  \bibinfo{journal}{Physical Review Letters} \textbf{\bibinfo{volume}{85}},
  \bibinfo{pages}{3777} (\bibinfo{year}{2000}).

\bibitem[{\citenamefont{Lundblad et~al.}(2008)\citenamefont{Lundblad, Lee,
  Spielman, Brown, Phillips, and Porto}}]{RefWorks:348}
\bibinfo{author}{\bibfnamefont{N.}~\bibnamefont{Lundblad}},
  \bibinfo{author}{\bibfnamefont{P.~J.} \bibnamefont{Lee}},
  \bibinfo{author}{\bibfnamefont{I.~B.} \bibnamefont{Spielman}},
  \bibinfo{author}{\bibfnamefont{B.~L.} \bibnamefont{Brown}},
  \bibinfo{author}{\bibfnamefont{W.~D.} \bibnamefont{Phillips}},
  \bibnamefont{and} \bibinfo{author}{\bibfnamefont{J.~V.} \bibnamefont{Porto}},
  \bibinfo{journal}{Physical Review Letters} \textbf{\bibinfo{volume}{100}},
  \bibinfo{pages}{150401} (\bibinfo{year}{2008}).

\bibitem[{\citenamefont{Hofferberth et~al.}(2006)\citenamefont{Hofferberth,
  Lesanovsky, Fischer, Verdu, and Schmiedmayer}}]{RefWorks:334}
\bibinfo{author}{\bibfnamefont{S.}~\bibnamefont{Hofferberth}},
  \bibinfo{author}{\bibfnamefont{I.}~\bibnamefont{Lesanovsky}},
  \bibinfo{author}{\bibfnamefont{B.}~\bibnamefont{Fischer}},
  \bibinfo{author}{\bibfnamefont{J.}~\bibnamefont{Verdu}}, \bibnamefont{and}
  \bibinfo{author}{\bibfnamefont{J.}~\bibnamefont{Schmiedmayer}},
  \bibinfo{journal}{Nature Physics} \textbf{\bibinfo{volume}{2}},
  \bibinfo{pages}{710} (\bibinfo{year}{2006}).

\bibitem[{\citenamefont{Colombe et~al.}(2004)\citenamefont{Colombe, Knyazchyan,
  Morizot, Mercier, Lorent, and Perrin}}]{RefWorks:336}
\bibinfo{author}{\bibfnamefont{Y.}~\bibnamefont{Colombe}},
  \bibinfo{author}{\bibfnamefont{E.}~\bibnamefont{Knyazchyan}},
  \bibinfo{author}{\bibfnamefont{O.}~\bibnamefont{Morizot}},
  \bibinfo{author}{\bibfnamefont{B.}~\bibnamefont{Mercier}},
  \bibinfo{author}{\bibfnamefont{V.}~\bibnamefont{Lorent}}, \bibnamefont{and}
  \bibinfo{author}{\bibfnamefont{H.}~\bibnamefont{Perrin}},
  \bibinfo{journal}{Europhysics Letters} \textbf{\bibinfo{volume}{67}},
  \bibinfo{pages}{593} (\bibinfo{year}{2004}).

\bibitem[{\citenamefont{Heathcote et~al.}(2008)\citenamefont{Heathcote, Nugent,
  Sheard, and Foot}}]{RefWorks:375}
\bibinfo{author}{\bibfnamefont{W.~H.} \bibnamefont{Heathcote}},
  \bibinfo{author}{\bibfnamefont{E.}~\bibnamefont{Nugent}},
  \bibinfo{author}{\bibfnamefont{B.~T.} \bibnamefont{Sheard}},
  \bibnamefont{and} \bibinfo{author}{\bibfnamefont{C.~J.} \bibnamefont{Foot}},
  \bibinfo{journal}{New Journal of Physics} \textbf{\bibinfo{volume}{10}},
  \bibinfo{pages}{043012} (\bibinfo{year}{2008}).

\bibitem[{\citenamefont{Yi et~al.}(2008)\citenamefont{Yi, Daley, Pupillo, and
  Zoller}}]{RefWorks:383}
\bibinfo{author}{\bibfnamefont{W.}~\bibnamefont{Yi}},
  \bibinfo{author}{\bibfnamefont{A.~J.} \bibnamefont{Daley}},
  \bibinfo{author}{\bibfnamefont{G.}~\bibnamefont{Pupillo}}, \bibnamefont{and}
  \bibinfo{author}{\bibfnamefont{P.}~\bibnamefont{Zoller}},
  \bibinfo{journal}{New Journal of Physics} \textbf{\bibinfo{volume}{10}},
  \bibinfo{pages}{073015} (\bibinfo{year}{2008}).

\bibitem[{\citenamefont{Tiesinga et~al.}(1993)\citenamefont{Tiesinga, Verhaar,
  and Stoof}}]{RefWorks:379}
\bibinfo{author}{\bibfnamefont{E.}~\bibnamefont{Tiesinga}},
  \bibinfo{author}{\bibfnamefont{B.~J.} \bibnamefont{Verhaar}},
  \bibnamefont{and} \bibinfo{author}{\bibfnamefont{H.~T.~C.}
  \bibnamefont{Stoof}}, \bibinfo{journal}{Physical Review A}
  \textbf{\bibinfo{volume}{47}}, \bibinfo{pages}{4114} (\bibinfo{year}{1993}).

\bibitem[{\citenamefont{Inouye et~al.}(1998)\citenamefont{Inouye, Andrews,
  Stenger, Miesner, Stamper-Kurn, and Ketterle}}]{RefWorks:378}
\bibinfo{author}{\bibfnamefont{S.}~\bibnamefont{Inouye}},
  \bibinfo{author}{\bibfnamefont{M.~R.} \bibnamefont{Andrews}},
  \bibinfo{author}{\bibfnamefont{J.}~\bibnamefont{Stenger}},
  \bibinfo{author}{\bibfnamefont{H.~J.} \bibnamefont{Miesner}},
  \bibinfo{author}{\bibfnamefont{D.~M.} \bibnamefont{Stamper-Kurn}},
  \bibnamefont{and} \bibinfo{author}{\bibfnamefont{W.}~\bibnamefont{Ketterle}},
  \bibinfo{journal}{Nature} \textbf{\bibinfo{volume}{392}},
  \bibinfo{pages}{151} (\bibinfo{year}{1998}).

\bibitem[{\citenamefont{Gupta et~al.}(1996)\citenamefont{Gupta, McClelland,
  Marte, and Celotta}}]{RefWorks:358}
\bibinfo{author}{\bibfnamefont{R.}~\bibnamefont{Gupta}},
  \bibinfo{author}{\bibfnamefont{J.~J.} \bibnamefont{McClelland}},
  \bibinfo{author}{\bibfnamefont{P.}~\bibnamefont{Marte}}, \bibnamefont{and}
  \bibinfo{author}{\bibfnamefont{R.~J.} \bibnamefont{Celotta}},
  \bibinfo{journal}{Physical Review Letters} \textbf{\bibinfo{volume}{76}},
  \bibinfo{pages}{4689} (\bibinfo{year}{1996}).

\bibitem[{\citenamefont{Zhang et~al.}(2005)\citenamefont{Zhang, Morrow, Berman,
  and Raithel}}]{RefWorks:361}
\bibinfo{author}{\bibfnamefont{R.}~\bibnamefont{Zhang}},
  \bibinfo{author}{\bibfnamefont{N.~V.} \bibnamefont{Morrow}},
  \bibinfo{author}{\bibfnamefont{P.~R.} \bibnamefont{Berman}},
  \bibnamefont{and} \bibinfo{author}{\bibfnamefont{G.}~\bibnamefont{Raithel}},
  \bibinfo{journal}{Physical Review A} \textbf{\bibinfo{volume}{72}},
  \bibinfo{pages}{043409} (\bibinfo{year}{2005}).

\bibitem[{\citenamefont{Lu et~al.}(2006)\citenamefont{Lu, Li, and
  Yu}}]{RefWorks:371}
\bibinfo{author}{\bibfnamefont{X.}~\bibnamefont{Lu}},
  \bibinfo{author}{\bibfnamefont{J.}~\bibnamefont{Li}}, \bibnamefont{and}
  \bibinfo{author}{\bibfnamefont{Y.}~\bibnamefont{Yu}},
  \bibinfo{journal}{Physical Review A} \textbf{\bibinfo{volume}{73}},
  \bibinfo{pages}{043607} (\bibinfo{year}{2006}).

\bibitem[{\citenamefont{Grimm et~al.}(2000)\citenamefont{Grimm, Weidemuller,
  and Ovchinnikov}}]{RefWorks:381}
\bibinfo{author}{\bibfnamefont{R.}~\bibnamefont{Grimm}},
  \bibinfo{author}{\bibfnamefont{M.}~\bibnamefont{Weidemuller}},
  \bibnamefont{and} \bibinfo{author}{\bibfnamefont{Y.~B.}
  \bibnamefont{Ovchinnikov}} (\bibinfo{year}{2000}), vol.~\bibinfo{volume}{42}
  of \emph{\bibinfo{series}{Advances In Atomic, Molecular, and Optical
  Physics}}, pp. \bibinfo{pages}{95--170}.

\bibitem[{\citenamefont{Cohen-Tannoudji
  et~al.}(2004)\citenamefont{Cohen-Tannoudji, Dupont-Roc, and
  Grynberg}}]{RefWorks:377}
\bibinfo{author}{\bibfnamefont{C.}~\bibnamefont{Cohen-Tannoudji}},
  \bibinfo{author}{\bibfnamefont{J.}~\bibnamefont{Dupont-Roc}},
  \bibnamefont{and} \bibinfo{author}{\bibfnamefont{G.}~\bibnamefont{Grynberg}},
  \emph{\bibinfo{title}{Atom-Photon Interactions}} (\bibinfo{year}{2004}), pp.
  \bibinfo{pages}{370--372}.

\bibitem[{\citenamefont{Marzari and Vanderbilt}(1997)}]{RefWorks:352}
\bibinfo{author}{\bibfnamefont{N.}~\bibnamefont{Marzari}} \bibnamefont{and}
  \bibinfo{author}{\bibfnamefont{D.}~\bibnamefont{Vanderbilt}},
  \bibinfo{journal}{Physical Review B} \textbf{\bibinfo{volume}{56}},
  \bibinfo{pages}{12847} (\bibinfo{year}{1997}).

\bibitem[{\citenamefont{Roberts et~al.}(1998)\citenamefont{Roberts, Claussen,
  Burke, Greene, Cornell, and Wieman}}]{RefWorks:380}
\bibinfo{author}{\bibfnamefont{J.~L.} \bibnamefont{Roberts}},
  \bibinfo{author}{\bibfnamefont{N.~R.} \bibnamefont{Claussen}},
  \bibinfo{author}{\bibfnamefont{J.~P.} \bibnamefont{Burke}},
  \bibinfo{author}{\bibfnamefont{C.~H.} \bibnamefont{Greene}},
  \bibinfo{author}{\bibfnamefont{E.~A.} \bibnamefont{Cornell}},
  \bibnamefont{and} \bibinfo{author}{\bibfnamefont{C.~E.}
  \bibnamefont{Wieman}}, \bibinfo{journal}{Physical Review Letters}
  \textbf{\bibinfo{volume}{81}}, \bibinfo{pages}{5109} (\bibinfo{year}{1998}).

\bibitem[{\citenamefont{LeBlanc and Thywissen}(2007)}]{RefWorks:320}
\bibinfo{author}{\bibfnamefont{L.~J.} \bibnamefont{LeBlanc}} \bibnamefont{and}
  \bibinfo{author}{\bibfnamefont{J.~H.} \bibnamefont{Thywissen}},
  \bibinfo{journal}{Physical Review A} \textbf{\bibinfo{volume}{75}},
  \bibinfo{pages}{053612} (\bibinfo{year}{2007}).

\bibitem[{\citenamefont{Zener}(1932)}]{RefWorks:384}
\bibinfo{author}{\bibfnamefont{C.}~\bibnamefont{Zener}},
  \bibinfo{journal}{Proceedings of the Royal Society (Series A)}
  \textbf{\bibinfo{volume}{137}}, \bibinfo{pages}{696} (\bibinfo{year}{1932}).

\bibitem[{\citenamefont{Moerdijk et~al.}(1996)\citenamefont{Moerdijk, Verhaar,
  and Nagtegaal}}]{RefWorks:333}
\bibinfo{author}{\bibfnamefont{A.~J.} \bibnamefont{Moerdijk}},
  \bibinfo{author}{\bibfnamefont{B.~J.} \bibnamefont{Verhaar}},
  \bibnamefont{and} \bibinfo{author}{\bibfnamefont{T.~M.}
  \bibnamefont{Nagtegaal}}, \bibinfo{journal}{Physical Review A}
  \textbf{\bibinfo{volume}{53}}, \bibinfo{pages}{4343} (\bibinfo{year}{1996}).

\bibitem[{\citenamefont{Suominen et~al.}(1998)\citenamefont{Suominen, Tiesinga,
  and Julienne}}]{RefWorks:337}
\bibinfo{author}{\bibfnamefont{K.~A.} \bibnamefont{Suominen}},
  \bibinfo{author}{\bibfnamefont{E.}~\bibnamefont{Tiesinga}}, \bibnamefont{and}
  \bibinfo{author}{\bibfnamefont{P.~S.} \bibnamefont{Julienne}},
  \bibinfo{journal}{Physical Review A} \textbf{\bibinfo{volume}{58}},
  \bibinfo{pages}{3983} (\bibinfo{year}{1998}).

\bibitem[{\citenamefont{Daley}()}]{RefWorks:385}
\bibinfo{author}{\bibfnamefont{A.~J.} \bibnamefont{Daley}},
  \bibinfo{note}{private communication}.

\bibitem[{\citenamefont{Brys et~al.}(2005)\citenamefont{Brys, Czekaj, Daum,
  Fierlinger, George, Henneck, Hochman, Kasprzak, Kohlik, Kirch
  et~al.}}]{RefWorks:446}
\bibinfo{author}{\bibfnamefont{T.}~\bibnamefont{Brys}},
  \bibinfo{author}{\bibfnamefont{S.}~\bibnamefont{Czekaj}},
  \bibinfo{author}{\bibfnamefont{M.}~\bibnamefont{Daum}},
  \bibinfo{author}{\bibfnamefont{P.}~\bibnamefont{Fierlinger}},
  \bibinfo{author}{\bibfnamefont{D.}~\bibnamefont{George}},
  \bibinfo{author}{\bibfnamefont{R.}~\bibnamefont{Henneck}},
  \bibinfo{author}{\bibfnamefont{Z.}~\bibnamefont{Hochman}},
  \bibinfo{author}{\bibfnamefont{M.}~\bibnamefont{Kasprzak}},
  \bibinfo{author}{\bibfnamefont{K.}~\bibnamefont{Kohlik}},
  \bibinfo{author}{\bibfnamefont{K.}~\bibnamefont{Kirch}},
  \bibnamefont{et~al.}, \bibinfo{journal}{Journal of Research of the National
  Institute of Standards and Technology} \textbf{\bibinfo{volume}{110}},
  \bibinfo{pages}{173} (\bibinfo{year}{2005}).

\bibitem[{\citenamefont{Öttl et~al.}(2006)\citenamefont{Öttl, Ritter,
  Köhl, and Esslinger}}]{RefWorks:445}
\bibinfo{author}{\bibfnamefont{A.}~\bibnamefont{Öttl}},
  \bibinfo{author}{\bibfnamefont{S.}~\bibnamefont{Ritter}},
  \bibinfo{author}{\bibfnamefont{M.}~\bibnamefont{Köhl}}, \bibnamefont{and}
  \bibinfo{author}{\bibfnamefont{T.}~\bibnamefont{Esslinger}},
  \bibinfo{journal}{Review of Scientific Instruments}
  \textbf{\bibinfo{volume}{77}} (\bibinfo{year}{2006}).

\end{thebibliography}

\end{document}